\documentclass[11pt]{article}
\usepackage{amsmath,amsfonts,amssymb,amsthm}
\usepackage{graphicx,setspace,array,algorithmic,algorithm,subfigure}
\usepackage[pdftex,colorlinks=true,linkcolor=blue,citecolor=blue,urlcolor=blue,bookmarks=false,pdfpagemode=None]{hyperref}
\usepackage{natbib}
 \bibpunct{(}{)}{;}{a}{,}{,}
\usepackage[top=0.75in, bottom=0.75in, left=0.75in, right=0.75in]{geometry}
\usepackage{url}
\usepackage{longtable}
\usepackage{hyphenat}
\usepackage{latexsym}
\usepackage{fullpage}

\newtheorem{theorem}{Theorem}

\author{Edoardo M. Airoldi, David S. Choi, and Patrick J. Wolfe\smallskip\\Harvard University, Cambridge, MA 02138, USA\smallskip\\{\small(airoldi@stat.harvard.edu, dchoi@seas.harvard.edu, patrick@seas.harvard.edu)}}
\title{Confidence sets for network structure}
\date{}
\begin{document}
\maketitle

\begin{abstract}
Latent variable models are frequently used to identify structure in dichotomous network data, in part because they give rise to a Bernoulli product likelihood that is both well understood and consistent with the notion of exchangeable random graphs.  In this article we propose conservative confidence sets that hold with respect to these underlying Bernoulli parameters as a function of any given partition of network nodes, enabling us to assess estimates of \emph{residual} network structure, that is, structure that cannot be explained by known covariates and thus cannot be easily verified by manual inspection. 
We demonstrate the proposed methodology by analyzing student friendship networks from the National Longitudinal Survey of Adolescent Health that include race, gender, and school year as covariates.  We employ a stochastic expectation-maximization algorithm to fit a logistic regression model that includes these explanatory variables as well as a latent stochastic blockmodel component and additional node-specific effects.  Although maximum-likelihood estimates do not appear consistent in this context, we are able to evaluate confidence sets as a function of different blockmodel partitions, which enables us to qualitatively assess the significance of estimated residual network structure relative to a baseline, which models covariates but lacks block structure.
\end{abstract}

\section{Introduction}

Network datasets comprising \emph{edge} measurements $A_{ij} \in \{0,1\}$ of a binary, symmetric, and anti-reflexive relation on a set of $n$ nodes, $1 \leq i < j \leq n$, are fast becoming of paramount interest in the statistical analysis and data mining literatures~\citep{Zhen:Gold:Fien:Airo:2009}.  A common aim of many models for such data is to test for and explain the presence of network \emph{structure}, primary examples being \emph{communities} and \emph{blocks} of nodes that are equivalent in some formal sense.  Algorithmic formulations of this problem take varied forms and span many literatures, touching on subjects as diverse as spectral graph theory~\citep{chun:2003}, statistical physics~\citep{Albe:Bara:2002,Newm:2003}, theoretical computer science~\citep{Coop:Frie:2003,achl:dsou:spen:2009,easl:klei:2010}, economics~\citep{jack:2008}, and social network analysis~\citep{Wass:Faus:1994}.

One popular modeling assumption for network data is to assume dyadic independence of the edge measurements when conditioned on a set of latent variables~\citep{Snij:Nowi:1997, handcock2007model, airoldi2008mixed, hoff2008modeling}.  The number of latent parameters in such models generally increases with the size of the graph, however, meaning that computationally intensive fitting algorithms may be required and standard consistency results may not always hold. As a result, it can often be difficult to assess statistical significance or quantify the uncertainty associated with parameter estimates. This issue is evident in literatures focused on community detection, where common practice is to examine whether algorithmically identified communities agree with prior knowledge or intuition~\citep{newman2006modularity, traud2010community}. This practice is less useful if additional confirmatory information is unavailable, or if detailed uncertainty quantification is desired.

Confidence sets are a standard statistical tool for uncertainty quantification, but they are not yet well developed for network data~\citep{Wass:Faus:1994}.
In this paper, we propose a family of confidence sets for network structure that apply under the assumption of a Bernoulli product likelihood. The form of these sets stems from a stochastic blockmodel formulation which reflects the notion of latent nodal classes, and they provide a new tool for the analysis of estimated or algorithmically determined network structure.  We demonstrate usage of the confidence sets by analyzing a sample of 26 adolescent friendship networks from the National Longitudinal Survey of Adolescent Health (available at http://www.cpc.unc.edu/addhealth), using a \emph{baseline}  model that only includes explanatory covariates and heterogeneity in the nodal degrees. We employ these confidence sets to validate departures from this baseline model taking the form of \emph{residual} community structure.  Though the confidence sets we employ are known to be conservative~\citep{mypaper}, we show that they are effective in identifying putative residual structure in these friendship network data.

\section{Model specification and inference}

We represent dichotomous network data via a sociomatrix $A \in \{0,1\}^{N \times N}$ that reflects the adjacency structure of a simple, undirected graph on $N$ nodes.  In keeping with the latent variable network analysis literature, we assume entries $\{A_{ij}\}$ for $i < j$ to be independent Bernoulli random variables with associated success probabilities $\{P_{ij}\}_{i < j}$, and complete $A$ as a symmetric matrix with zeros along its main diagonal.  The corresponding data log-likelihood is given by
\begin{equation}\label{eq:logLik}
L(A;P) = \sum_{i<j} A_{ij} \log(P_{ij}) + (1-A_{ij})\log(1-P_{ij}),
\end{equation}
where each $P_{ij}$ can itself be modeled as a function of latent as well as explanatory variables.  Typically, the nodal indices themselves impart no inferential information and the latent variable formulation is exchangeable; in fact, in lieu of independence one may start with a much weaker assumption that the graph itself is generated from an exchangeable distribution (see~\citet{bickel2009nonparametric, hoff2009multiplicative}).  Exchangeability accommodates many types of dependencies that are of interest in network analysis, such as triadic closure.

Given an instantiation of $A$ and a latent variable model for the probabilities $\{P_{ij}\}_{i < j}$, it is natural to seek a quantification of the uncertainty associated with estimates of these Bernoulli parameters.  A standard approach in non-graphical settings is to posit a parametric form for each $P_{ij}$ and then compute confidence intervals with respect to the corresponding parameter values, for example by appealing to standard maximum-likelihood asymptotics.  However, as mentioned earlier, the formulation of most latent variable network models dictates an increasing number of parameters as the number of network nodes grows; as we have observed in our own exploratory data analysis, this amount of expressive power appears necessary to capture many salient characteristics of network data.  As a result, standard asymptotic results do not necessarily apply, leaving open questions for inference.

\subsection{A logistic regression model for network structure}

To illustrate the complexities that can quickly arise in this inferential setting, we adopt a latent variable network model with a standard flavor: a logistic regression model that simultaneously incorporates aspects of blockmodels, additive effects, and explanatory variables (see~\citet{hoff2009multiplicative} for a more general formulation).  Specifically, we incorporate a $K$-class stochastic blockmodel component parameterized in terms of a symmetric matrix $\Theta \in \mathbb{R}^{K\times K}$ and a membership vector $z \in \{1,\ldots,K\}^N$ whose values denote the class of each node, with $P_{ij}$ depending on $\Theta_{z_i z_j}$.  A vector of additional node-specific latent variables $\alpha$ is included to account for heterogeneity in the observed nodal degrees, along with a vector of regression coefficients $\beta$ corresponding to explanatory variables $x(i,j)$.  Thus we obtain the log-odds parameterization
\begin{equation}\label{eq:logodds}
\log \frac{P_{ij}}{1-P_{ij}} = \Theta_{z_i z_j} + \alpha_i + \alpha_j + x(i,j)' \beta,
\end{equation}
where we further enforce the identifiability constraint that $\sum_i \alpha_i = 0$.

Intuitively, the blockmodel component of model~\eqref{eq:logodds} serves to enforce stochastic equivalence of nodes in the same class~\citep{holland1983stochastic, fienberg1985statistical}, after accounting for node-specific additive effects and covariates.  Blockmodeling in this manner is historically motivated by the deterministic notion of structural equivalence~\citep{lorrain1971structural, doreian2005generalized}, under which the network must remain identical under any relabeling of nodes within each community.  While this notion is typically too strong to hold exactly, relaxations such as regular equivalence\footnote{In deterministic equivalence, the bipartite subgraph of edges between two communities is either empty or complete. In regular equivalence, the same subgraph is either empty, or has no isolated nodes. See~\citep{doreian2005generalized} for further discussion.} and other variations \citep{doreian2005generalized} can be thought of as deterministic precursors to the stochastic blockmodel~\citep{nowicki2001estimation} and its modern extensions~\citep{airoldi2005latent, airoldi2008mixed, xing:fu:song:2010}.

\subsection{Likelihood-based inference}

Analogous to the deterministic setting, exact maximization of the corresponding log-likelihood $L(A; z,\Theta,\alpha,\beta, x)$ is computationally demanding even for moderately large $K$ and $N$, owing to the total number of possible nodal partitions induced by $z$. However, when $z$ is fixed the corresponding conditional likelihood maximization task is convex, and so we may adopt a stochastic expectation-maximization (EM) approach to inference in which $z$ is treated as missing data~\citep{Demp:Lair:Rubi:1977,vand:meng:2010}.
\begin{algorithm}
\caption{\label{alg:1}Stochastic Expectation-Maximization Fitting of model~\eqref{eq:logodds}}
\begin{enumerate}
\item Set $t = 0$ and initialize $(z^{(0)},\Theta^{(0)},\alpha^{(0)},\beta^{(0)})$.
\item For iteration $t$, do:
\begin{description}
\item[E-step] Sample $z^{(t)} \propto \exp\{ L(z\,\vert\, A; \Theta^{(t)},\alpha^{(t)},\beta^{(t)}, x) \}$ \\(e.g., via Gibbs sampling)
\item[M-step] Set $(\Theta^{(t)},\alpha^{(t)},\beta^{(t)}) = \operatorname{argmax}_{\Theta,\alpha,\beta} L(\Theta,\alpha,\beta \,\vert\, z^{(t)}; A, x)$ \\(convex optimization)
\end{description}
\item Set $t \leftarrow t + 1$ and return to Step 2.
\end{enumerate}
\vspace{-0.75\baselineskip}%
\end{algorithm}

Exact sampling of $z$ according to the E-step of Algorithm~\ref{alg:1} requires computing the normalizing constant of $\exp\{ L(z\,\vert\, A; \Theta ,\alpha,\beta, x) \}$, and so we may instead approximate this step by a series of univariate draws from a coordinate-wise Gibbs sampler that conditions on all other elements of $z$ in turn.  When $K$ is moderate, both steps of Algorithm~\ref{alg:1} are computationally tractable, since the Gibbs sampler requires only discrete distributions over $K$ values, and the M-step optimization problem is convex. We note that the confidence sets developed in Section \ref{sec:confSets} will accommodate the sub-optimality of Algorithm~\ref{alg:1}.

When $\alpha$ and $\beta$ are fixed to zero, model~\eqref{eq:logodds} reduces to a re-parameterization of the standard stochastic blockmodel. Consistency results for this model have been developed for a range of conditions~\citep{Snij:Nowi:1997, nowicki2001estimation, bickel2009nonparametric, rohe2010spectral, mypaper, celisse2011consistency}; alternative fitting methods with lower computational complexity, such as spectral clustering~\citep{rohe2010spectral} and modularity maximization~\citep{bickel2009nonparametric}, are also available for the stochastic blockmodel, with~\citep{bickel2009nonparametric} noting that such approaches appear to perform well in practice. However, it is not clear how uncertainty in $z$ and $\Theta$ should be quantified or even concisely expressed: in this vein, previous efforts to assess the robustness of fitted structure include~\citep{karrer2008robustness}, in which community partitions are analyzed under perturbations of the network, and \citep{massen2006thermodynamics}, in which the behavior of local minima resulting from simulated annealing is examined; a likelihood-based test is proposed in~\citep{copic2009identifying} to compare sets of community divisions.

Without the blockmodel components $z$ and $\Theta$, the model of Eq.~\eqref{eq:logodds} reduces to a generalized linear model whose likelihood can be maximized by standard methods. If $\alpha$ is further constrained to equal $0$, the model is finite dimensional and standard asymptotic results for inference can be applied. Otherwise, the increasing dimensionality of $\alpha$ brings consistency into question, and in fact certain negative results are known for the related $p_1$ exponential random graph model~\citep{holland1981exponential}.  Specifically,~\citep{haberman1981comment} reports that the maximum likelihood estimator for the $p_1$ model exhibits bias with magnitude equal to its variance.  Although estimation error does converge asymptotically to zero for the $p_1$ model, it is generally not known how to generate confidence intervals or hypothesis tests;~\citep{wasserman1985statistical} prescribes reporting standard errors only as summary statistics, with no association to $p$-values.  The predictions of~\citep{haberman1981comment} were replicated (reported below) when fitting simulated data drawn from the model of Eq.~\eqref{eq:logodds} with parameters matched to observed characteristics of the Adolescent Health friendship networks.

\subsection{Confidence sets for network structure}
\label{sec:confSets}

Instead of quantifying the uncertainty associated with estimates of the model parameters $(z,\Theta,\alpha,\beta)$, we directly find confidence sets for the Bernoulli likelihood parameters $\{P_{ij}\}_{i<j}$.  To this end, for any fixed $K$ and class assignment $z$, define symmetric matrices $\bar{\Phi}, \hat{\Phi}$ in $[0,1]^{K \times K}$ element-wise for $1 \leq a \leq b$ as
\begin{equation*}
\bar{\Phi}_{ab}^{(z)} = \frac{1}{n_{ab}}\sum_{i < j} P_{ij} \, 1\{z_i=a,z_j=b\}, \quad
\hat{\Phi}_{ab}^{(z)} = \frac{1}{n_{ab}}\sum_{i<j} A_{ij} \, 1\{z_i=a,z_j=b\},
\end{equation*}
with $n_{ab}$ denoting the maximum number of possible edges between classes $a$ and $b$ (i.e., the corresponding number of Bernoulli trials).  Thus $\bar{\Phi}_{ab}^{(z)}$ is the expected proportion of edges between (or within, if $a=b$) classes $a$ and $b$, under class assignment $z$, and $\hat{\Phi}_{ab}^{(z)}$ is its corresponding sample proportion estimator.

Intuitively, $\bar{\Phi}^{(z)}$ measures \emph{assortativity} by $z$; whenever the sociomatrix $A$ is unstructured, elements of $\bar{\Phi}^{(z)}$ should be nearly uniform for any choice of partition $z$.  When strong community structure is present in $A$, however, these elements should instead be well separated for corresponding values of $z$.  Thus, it is of interest to examine a confidence set that relates $\hat{\Phi}_{ab}^{(z)}$ to its expected value $\bar{\Phi}^{(z)}$ for a range of partitions $z$.  To this end, we may define such a set by considering a weighted sum of the form $\sum_{a\leq b} n_{ab} D(\hat{\Phi}_{ab}^{(z)}||\bar{\Phi}_{ab}^{(z)})$, where
$D(p||p') = p\log(p/p') + (1-p) \log[(1-p)/(1-p')]$ denotes the (nonnegative) Kullback--Leibler divergence of a Bernoulli random variable with parameter $p'$ from that of one with parameter $p$.

A confidence set is then obtainable via direct application of the following theorem.
\begin{theorem}[Choi et al., 2011] 
Let $\{A_{ij}\}_{i <j}$ be comprised of $\binom{N}{2}$ independent $\operatorname{Bernoulli}(P_{ij})$ trials, and let $\mathcal{Z} = \{1,\ldots,K\}^N$. Then with probability at least $1-\delta$,
\begin{equation}\label{eq:bound}
\sup_{z \in \mathcal{Z}} \sum_{a\leq b} n_{ab} D(\hat{\Phi}_{ab}^{(z)} || \bar{\Phi}_{ab}^{(z)}) \leq N \log K + (K^2+K)\log\left(\frac{N}{K}+1\right) + \log\frac{1}{\delta}.
\end{equation}
\end{theorem}
Because Eq.~\eqref{eq:bound} holds uniformly over all class assignments, we may choose to apply it directly to the value of $z$ obtained from Algorithm~\ref{alg:1}---and because it does not assume any particular form of latent structure, we are able to avoid the difficulties associated with determining confidence sets directly for the parameters of latent variable models such as Eq.~\eqref{eq:logodds}.  However, it is important to note that this generality comes at a price:  In simulation studies undertaken in~\citep{mypaper} as well as those detailed below, we have observed the bound of Eq.~\eqref{eq:bound} to be loose by a multiplicative factor ranging from 3 to 7 on average.

\subsection{Estimator consistency and confidence sets}
\label{sec:sim}

Recalling our above discussion of estimator consistency for the related $p_1$ model, we undertook a small simulation study to investigate the consistency of maximum-likelihood (ML) estimation in a ``baseline'' version of model~\eqref{eq:logodds} with $K=1$ and the corresponding (scalar) value of $\Theta$ set equal to zero.  We compared estimates for the cases $\alpha=0$ versus $\alpha$ unconstrained for 500 graphs generated randomly from a model of the form specified in Eq.~\eqref{eq:logodds} based on school 8 of the Add-Health data set. The number of nodes $N=204$ and covariates $x(i,j)$ matched that of School 8 in the Adolescent Health friendship network dataset, and the regression coefficient vector $\beta = (-2.6, 0.025, 0.9, -1.6)'$, set to match the ML estimate of $\beta$ for School 8, fitted via logistic regression with $\Theta=0, \alpha=0$. The covariates $x(i,j)$ comprised of an intercept term, an indicator for whether students $i$ and $j$ shared the same gender, an indicator for shared race, and their 
difference in school grade.

The inclusion of $\alpha$ in the model of Eq.~\eqref{eq:logodds} appears to give rise to a loss of estimator consistency, as shown in Table \ref{tab:sim} where the empirical bias of each component of $\beta$ is reported.
\begin{table}
\begin{center}
\begin{tabular}{>{\footnotesize}l|>{\footnotesize}r|>{\footnotesize}r|}
Element of $\beta$ & $\Theta=0, \alpha=0$ & $\Theta=0$ only \\
\hline
Intercept & $-$0.001 (0.004) & 2.26 (0.070) \\
Gender & 0.003 (0.004) & $-$0.005 (0.004) \\
Race & $-$0.001 (0.004) & $-$0.03 (0.005) \\
Grade & 0.006 (0.003) & 0.04 (0.003) \\
 \end{tabular}
       \caption{Empirical bias (with standard errors) of ML-estimated components of $\beta$ under a baseline model, for the cases $\alpha=0$ versus $\alpha$ unconstrained.  Note the change in estimated bias when $\alpha$ is included in the model.}
\label{tab:sim}
\end{center}
\end{table}
This suggests, as we alluded to above, that inferential conclusions based on parameter estimates from latent variable models should be interpreted with caution.  

\begin{figure}
\begin{center}
\includegraphics[width=0.6\columnwidth]{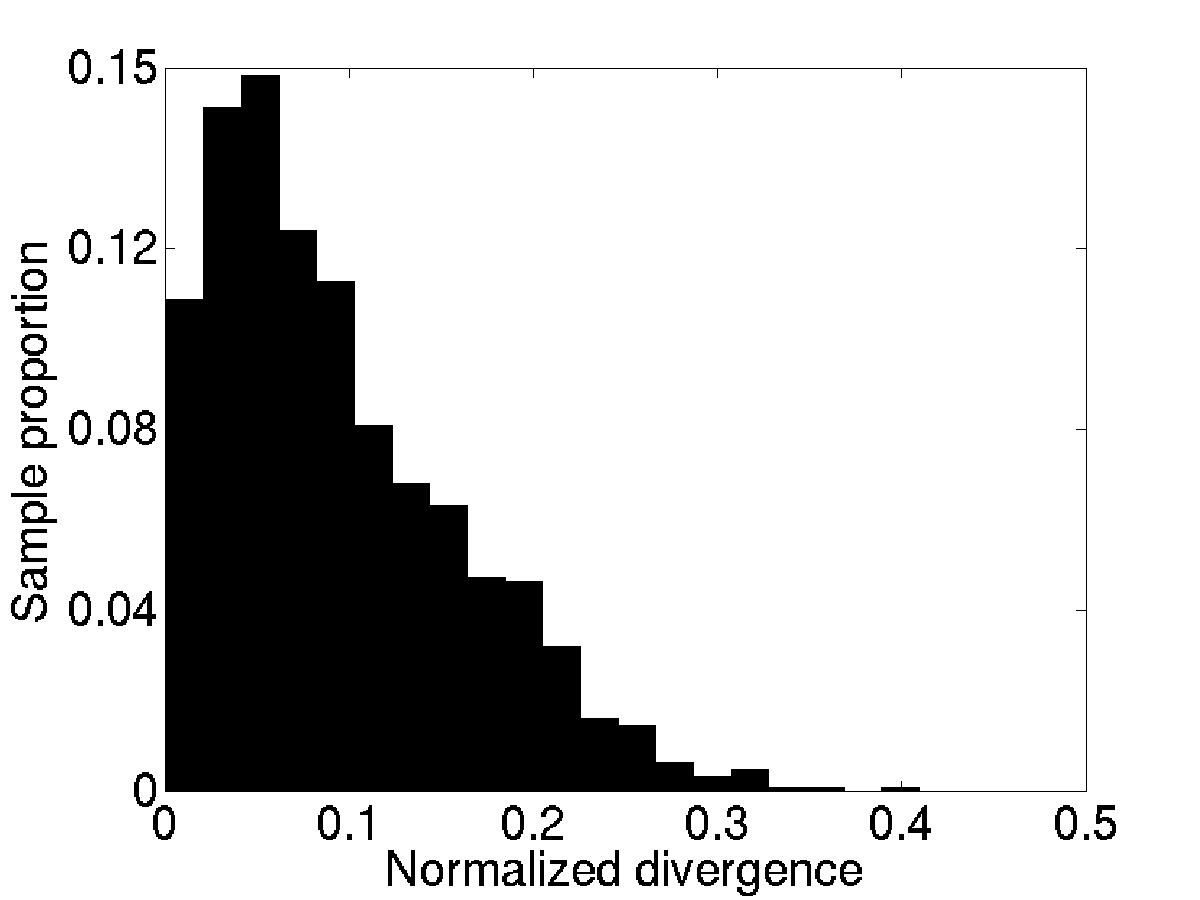}
\end{center}
\caption{Divergence terms $\sum_{a\leq b} n_{ab} D(\hat{\Phi}_{ab}^{(z)} || \bar{\Phi}_{ab}^{(z)})$ as fractions of 95\% confidence set values, shown for approximate maximum-likelihood fits to 1300 randomly graphs matched to the 26-school friendship network dataset. Smaller values indicate slack in the upper bound of Eq.~\eqref{eq:bound}.}
\label{fig:divergence histogram}
\end{figure}

To explore the tightness of the confidence sets given by the bound in Eq.~\eqref{eq:bound}, we fitted the full model specified in Eq.~\eqref{eq:logodds} with $K$ in the range 2--6 to 50 draws from a restricted version of the model corresponding to each of the 26 schools in our dataset.  In the same manner described above, each simulated graph shared the same size and covariates as its corresponding school in the dataset, with $\beta$ fixed to its ML-fitted value with $\Theta=0, \alpha=0$.  The empirical divergence term $\sum_{a\leq b} n_{ab} D(\hat{\Phi}_{ab}^{(z)} || \bar{\Phi}_{ab}^{(z)})$ under the approximate ML partition determined via Algorithm~\ref{alg:1} was then tabulated for each of these 1300 fits, and compared to its 95\% confidence upper bound given by Eq.~\eqref{eq:bound}. The empirical divergences are reported in the histogram of Fig.~\ref{fig:divergence histogram} as a fraction of the upper bound.

It may be seen from Fig.~\ref{fig:divergence histogram} that the largest divergence observed was less than 41\% of its corresponding bound, with 95\% of all divergences less than 22\% of their corresponding bound.  This analysis verifies that, despite the issues associated with estimator consistency in the context of latent variable models, our confidence sets can be used as a tool to assess putative network structure.  This analysis also provides an indication of how inflated the confidence set sizes are expected to be in practice; while conservative in nature, they seem usable for practical situations.

\section{Analysis of adolescent health networks}

We now turn our attention to an analysis of friendship networks from the National Longitudinal Survey of Adolescent Health; see, e.g.,~\citep{goodreau2009birds} for a recent analysis and in-depth discussion. Previous studies have looked for trends which held across several schools, such as clustering by race \citep{gonzalez2007community}. Within a school, however, it may be desirable to go beyond reported covariates such as gender, race, or grade in describing social structure. Here we examine the schools individually to find residual block structure not explained by the covariates. Since we will be unable to verify such blocks by checking against explanatory variables, we rely on the confidence sets developed above to assess significance of the discovered block structure. 

Recall from Section~\ref{sec:confSets} that Eq.~\eqref{eq:bound} enables us to calculate confidence sets with respect to Bernoulli parameters $\{P_{ij}\}$ for any class membership vector $z$ in terms of the corresponding sample proportion matrices $\hat{\Phi}^{(z)}$. Then, by comparing values of $\hat{\Phi}^{(z)}$ to a baseline model obtained by fitting $K=1,\Theta=0$ (thus removing the stochastic block component from Eq.~\eqref{eq:logodds}), we may evaluate whether or not the observed sample counts are consistent with the structure predicted by the baseline model.  This procedure provides a kind of notional $p$-value to qualitatively assess significance of the residual structure induced by any choice of $z$.

In Section~\ref{sec:dataDescription} below, we describe the dataset. Then, in Section~\ref{sec:modelChecking}, we verify empirically that the models  provide reasonable fits and conclusions, first by fitting our baseline model ($K=1$) and checking its behavior, and then by fitting a ``pure'' stochastic blockmodel ($K>1,\alpha=0,\beta=0$) and observing that we can recover block structure corresponding to held-out grade covariates. In Section~\ref{sec:residBlockStruct}, we consider estimates of $z$ obtained via Algorithm~\ref{alg:1} for $K$ in the range 2--6, and use our notion of confidence sets introduced above to evaluate how unlikely the corresponding observed configurations of $\hat{\Phi}^{(z)}$ are, relative to an underlying set of Bernoulli random variables whose parameters are assumed to be equal to those fitted under our baseline model.

\subsection{Network data description}
\label{sec:dataDescription}

The data we examine were collected in 1994--95 through an in-school survey administered to 90,118 students at 145 schools.   In the 26-school subset of the data we consider, each student was given a paper-and-pencil questionnaire and a copy of a roster listing every student in that school; the schools that we considered did not have a \emph{sister school} from where friends could also be listed (which would complicate the analysis) and they had fewer than 700 students. Students were asked to list up to five friends of each gender, and whether or not they had interacted with these friends at particular venues within a certain time period.  Gender, race (5 categories), and school year (grades 7--12) covariates were also recorded for each student. The number of students and reported friendships for this subset of schools are listed in Table~\ref{tab:school-stat}.

\begin{table}
\begin{center}
\begin{tabular}{>{\footnotesize}r|>{\footnotesize}r|>{\footnotesize}r||>{\footnotesize}r|>{\footnotesize}r|>{\footnotesize}r||>{\footnotesize}r|>{\footnotesize}r|>{\footnotesize}r|}
School & Nodes & Edges & School & Nodes & Edges & School & Nodes & Edges \\
\hline
1 & 69 & 220 & 21 & 377 & 1531 & 67 & 456 & 926 \\
2 & 105 & 349 & 22 & 614 & 2450 & 70 & 76 & 344\\
3 & 32 & 91 & 26 & 551 & 2066 & 71 & 74 & 358\\
5 & 157 & 730 & 29 & 569 & 2534 & 72 & 352 & 1398\\
6 & 108 & 378 & 38 & 521 & 1925 & 76 & 43 & 185\\
8 & 204 & 809 & 55 & 336 & 803 & 77 & 25 & 106\\
9 & 248 & 1004 & 56 & 446 & 1394 & 78 & 432 & 1334\\
10 & 678 & 2795 & 63 & 98 & 283 & 80 & 594 & 1745\\
18 & 284 & 1189 & 66 & 644 & 2865\\
\end{tabular}
\caption{Network characteristics (number of students and reported friendship edges) corresponding to each of the 26 schools used for analysis.}
\label{tab:school-stat}
\end{center}
\end{table}

We treat the friendship network data from these 26 schools as undirected graphs, with an edge present between two students whenever either of them indicated any type of friendship with the other. Other approaches are also possible and can be expected to influence inferential results~\citep{butts2003network}, though we verified through exploratory data analysis that estimates were qualitatively similar even if sparser versions of these networks were constructed by instead requiring mutual indication of friendship within student pairs, resulting in a roughly 3-fold decrease in the observed number of edges in our data (cf.~Table~\ref{tab:school-stat}).  As in Section \ref{sec:sim}, for each student pair the corresponding vector $x(i,j)$ of explanatory variables comprised an intercept term, binary indicator variables indicating shared gender and shared race, and an ordinal value equal to the absolute difference in the grades of the students; model~\eqref{eq:logodds} was then fitted to these data, with results described below.

\subsection{Model checking}
\label{sec:modelChecking}

We first fit model~\eqref{eq:logodds} with $\Theta=0$ and $\alpha=0$, since it reduces to a logistic regression with explanatory variables $x(i,j)$, for which standard asymptotic results apply. The parameter fits were examined and an analysis of deviance was conducted. The fits were observed to be well behaved in this setting; estimates of $\beta$ and their corresponding standard errors indicate a clustering effect by grade that is stronger than that of either shared gender or race.  The fitting thus appears consistent with the previous analysis of~\citep{handcock2007model}, in which student friendship clusters were observed to closely align with grades. An analysis of deviance, where each variable was withheld from the model, resulted in similar conclusions:  Average deviances across the 26 schools were $-$69, $-$238, and $-$3760 for gender, race, and grade respectively, with $p$-values below $0.05$ indicating significance in all but 3, 7, and 0 of the schools for each of the respective covariates; these schools had small numbers of students, with a maximum $N$ of 108.

Because standard analysis techniques do not apply to the parameters $\alpha$ and $\Theta$, before fitting the full model we first fit models using these terms separately, and checked if the results were intuitive. When $\alpha$ was re-introduced into the model of Eq.~\eqref{eq:logodds}, its components were observed to correlate highly with the sequence of observed nodal degrees in the data, as expected.  (Recall that consistency results are not known for this model, so that $p$-values cannot be associated with deviances or standard errors; however, in our simulations the maximum-likelihood estimates showed moderate errors, as discussed in Section~\ref{sec:sim}.)  For two of the schools, the resulting model was degenerate, whereas for the remaining schools the $\alpha$-degree correlation had a range of 0.78--0.94 and a median value of 0.89.  Estimates of $\beta$ did not undergo qualitative significant changes from their earlier values when the restriction $\alpha=0$ was lifted.

\begin{figure}[h!]
\subfigure[$K=2$]{
\makebox{\includegraphics[width=.3\linewidth]{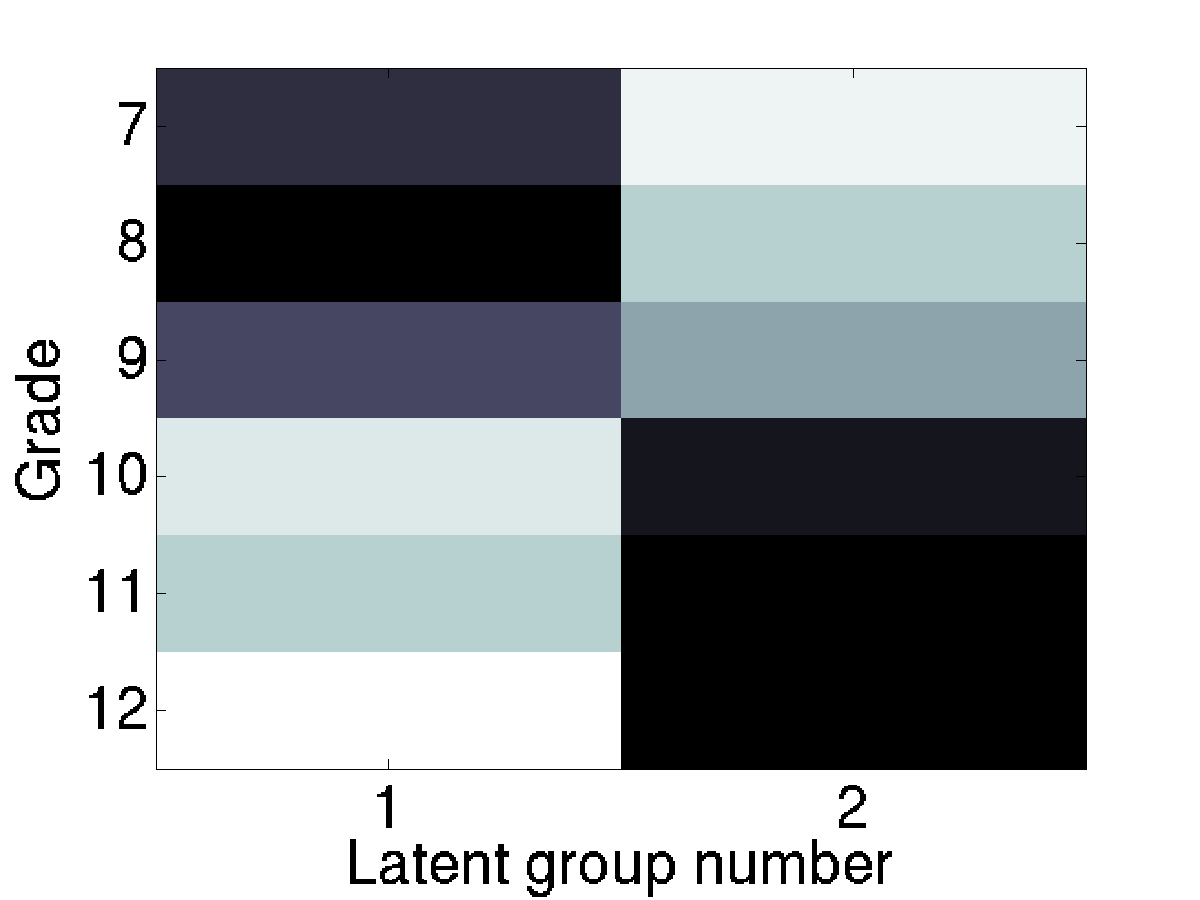}}
}
\subfigure[$K=3$]{
\makebox{\includegraphics[width=.3\linewidth]{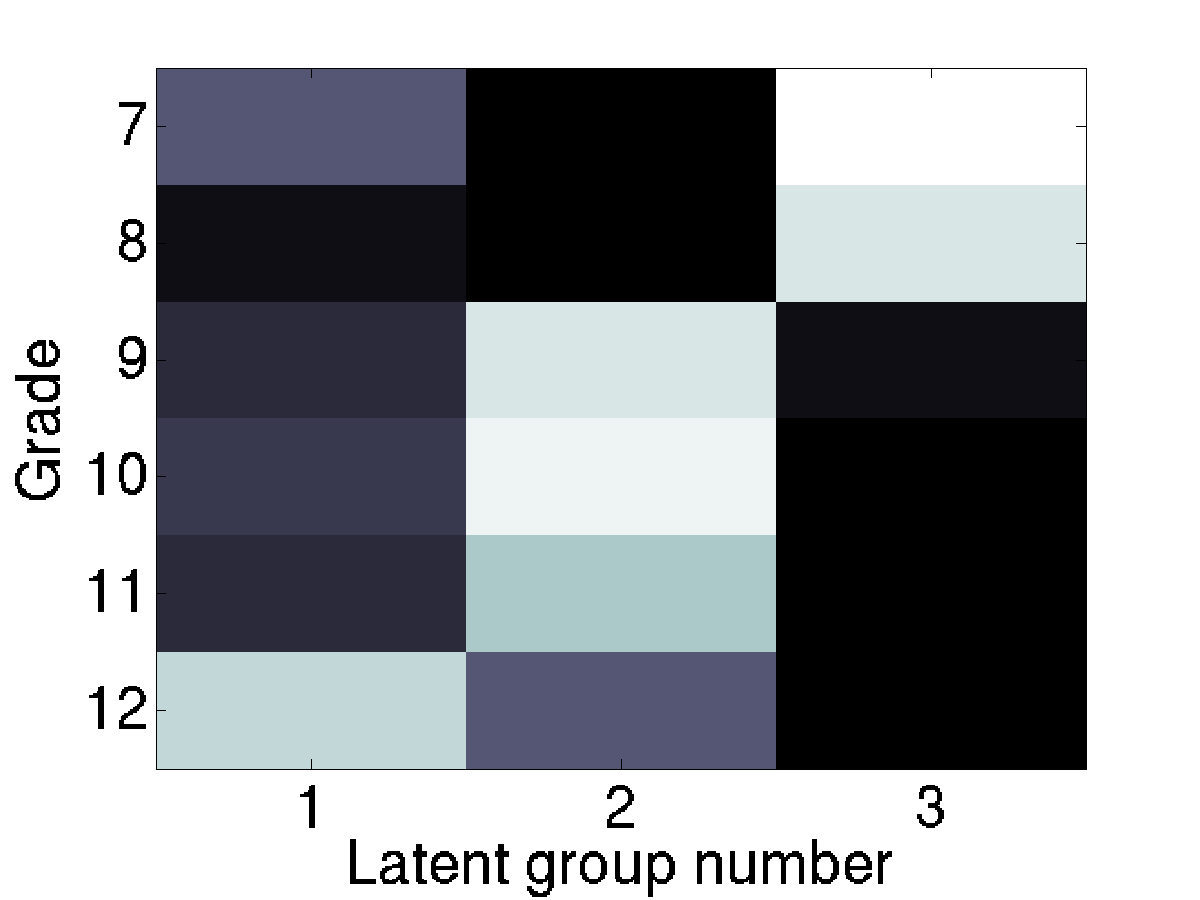}}
}
\subfigure[$K=4$]{
\makebox{\includegraphics[width=.3\linewidth]{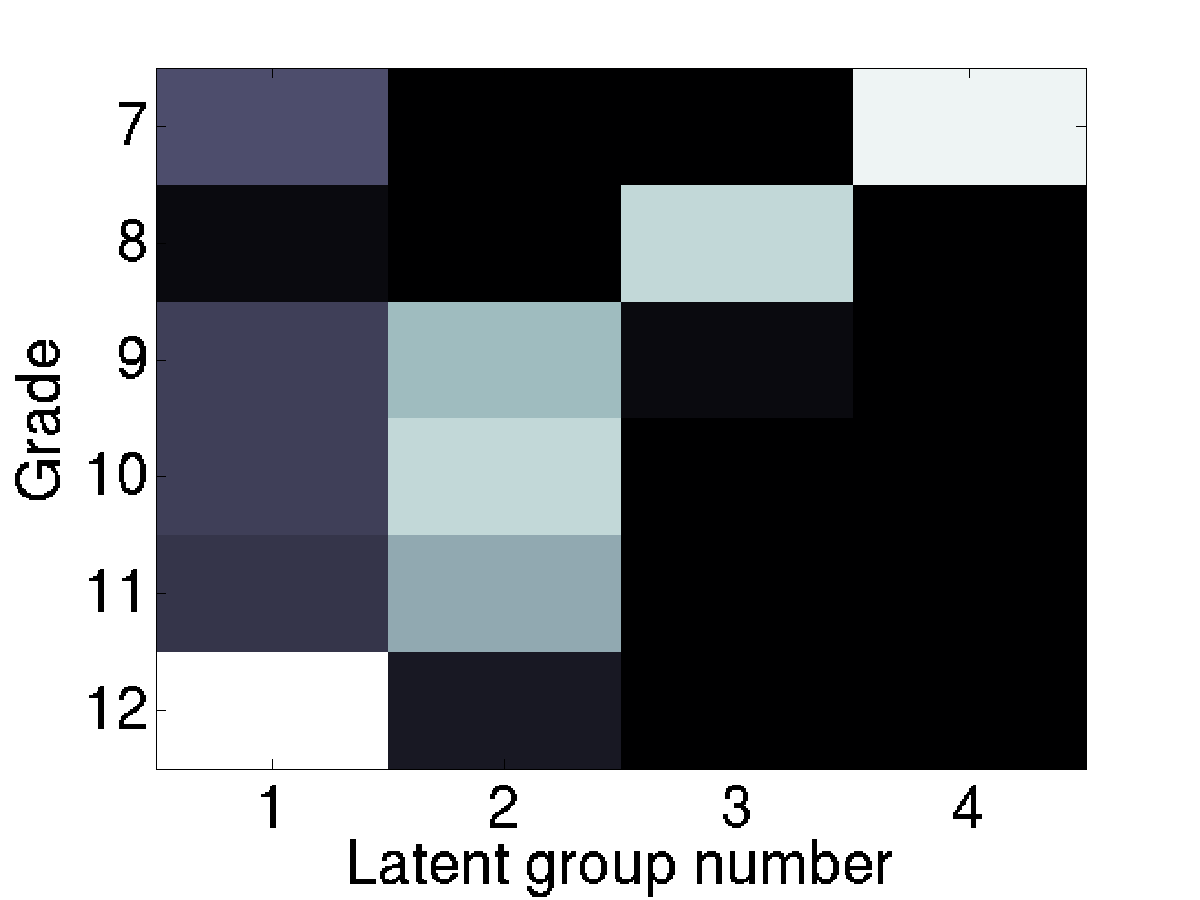}}
}
\subfigure[$K=5$]{
\makebox{\includegraphics[width=.3\linewidth]{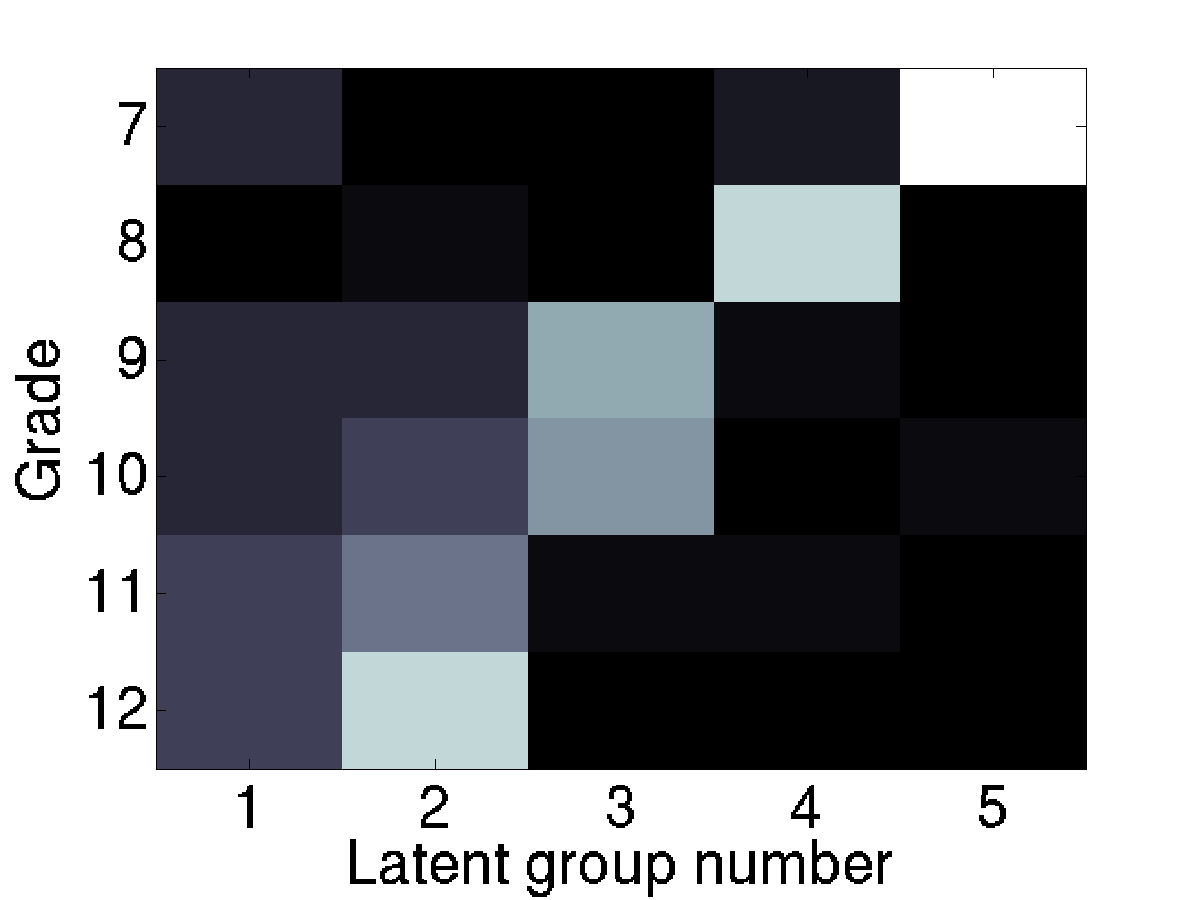}}
}
\subfigure[$K=6$]{
\makebox{\includegraphics[width=.3\linewidth]{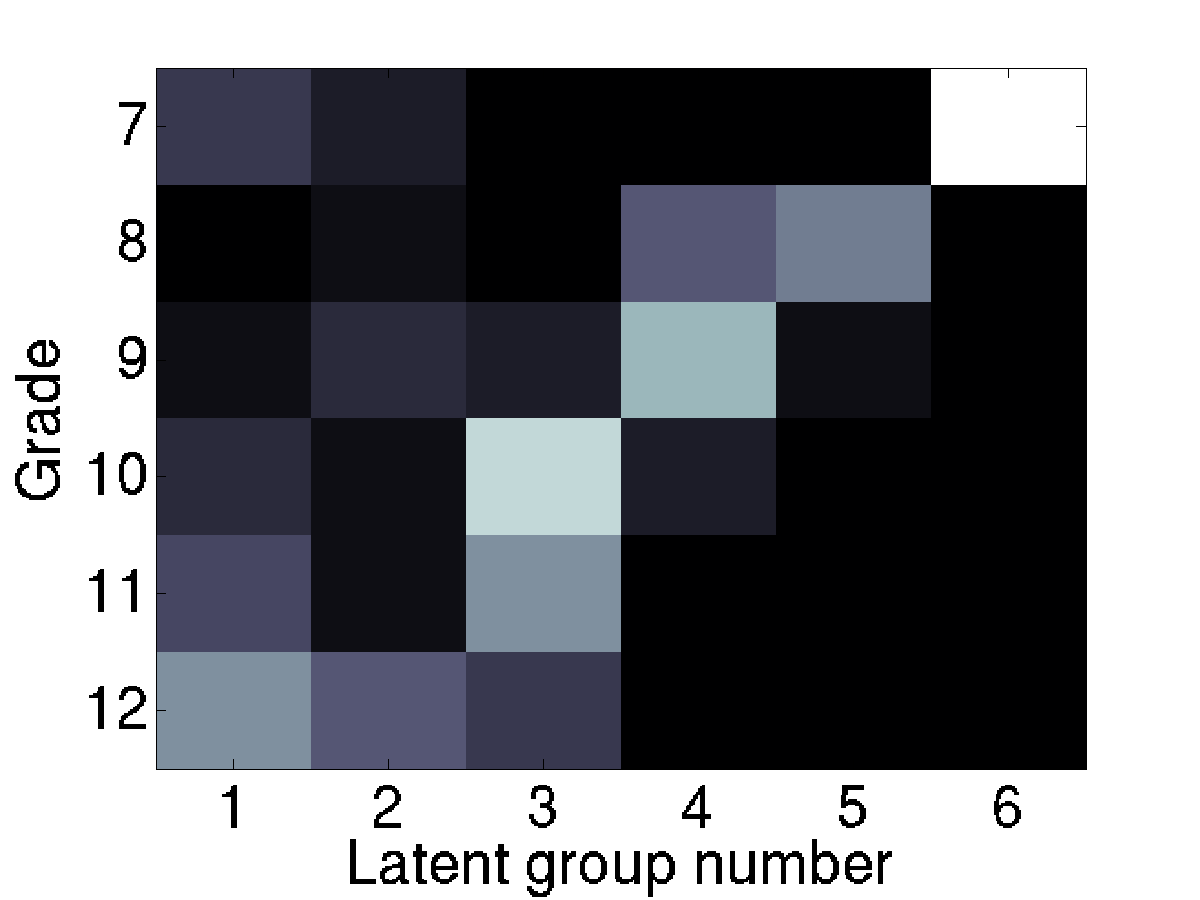}}
}
\caption{Student counts resulting from a stochastic blockmodel fit for $K \in \{2,\ldots,6\}$, arranged by latent block and school year (grade) for School 6.  The inferred block structure approximately aligns with the grade covariate (which was not included in this model).}
\label{fig: withheld covariate}
\end{figure}

A ``pure'' stochastic blockmodel ($\alpha=0, \beta=0$) was fitted to our data over the range $K \in \{2,\ldots,6\}$, to observe if the resulting block structure replicates that of any of the known covariates. Figure~\ref{fig: withheld covariate} shows counts of students by latent class (under the approximate maximum-likelihood estimate of $z$) and grade for School 6; it can be seen that the recovered grouping of students by latent class is closely aligned with the grade covariate, particularly for grades 7--10.

\subsection{Residual block structure}
\label{sec:residBlockStruct}

We now report on the assessment of residual block structure in the Adolescent Health friendship network data.  Recalling that the confidence sets obtained with Eq.~\eqref{eq:bound} hold uniformly for all partitions of equal size, independently of how they are chosen, we therefore may freely modify the fitting procedure of Algorithm~\ref{alg:1} to obtain partitions that exhibit the greatest degree of structure. Bearing in mind the high observed $\alpha$-degree correlation as discussed above, we replaced the latent variable vector $\alpha$ in the model of Eq.~\eqref{eq:logodds} with a more parsimonious categorical covariate determined by grouping the observed network degrees according to the ranges 0--3, 4--7, and 8--$\infty$. We also expanded the covariates by giving each race and grade pairing its own indicator function. These modifications would be inappropriate for the baseline model, as dyadic independence conditioned on the covariates would be lost, and standard errors for $\beta$ would be larger; however, the changes were useful for improving the output of Algorithm~\ref{alg:1} without invalidating Eq.~\eqref{eq:bound}.

\begin{figure}
\subfigure[School 10, $K=6$]{
\makebox{\includegraphics[width=.23\linewidth]{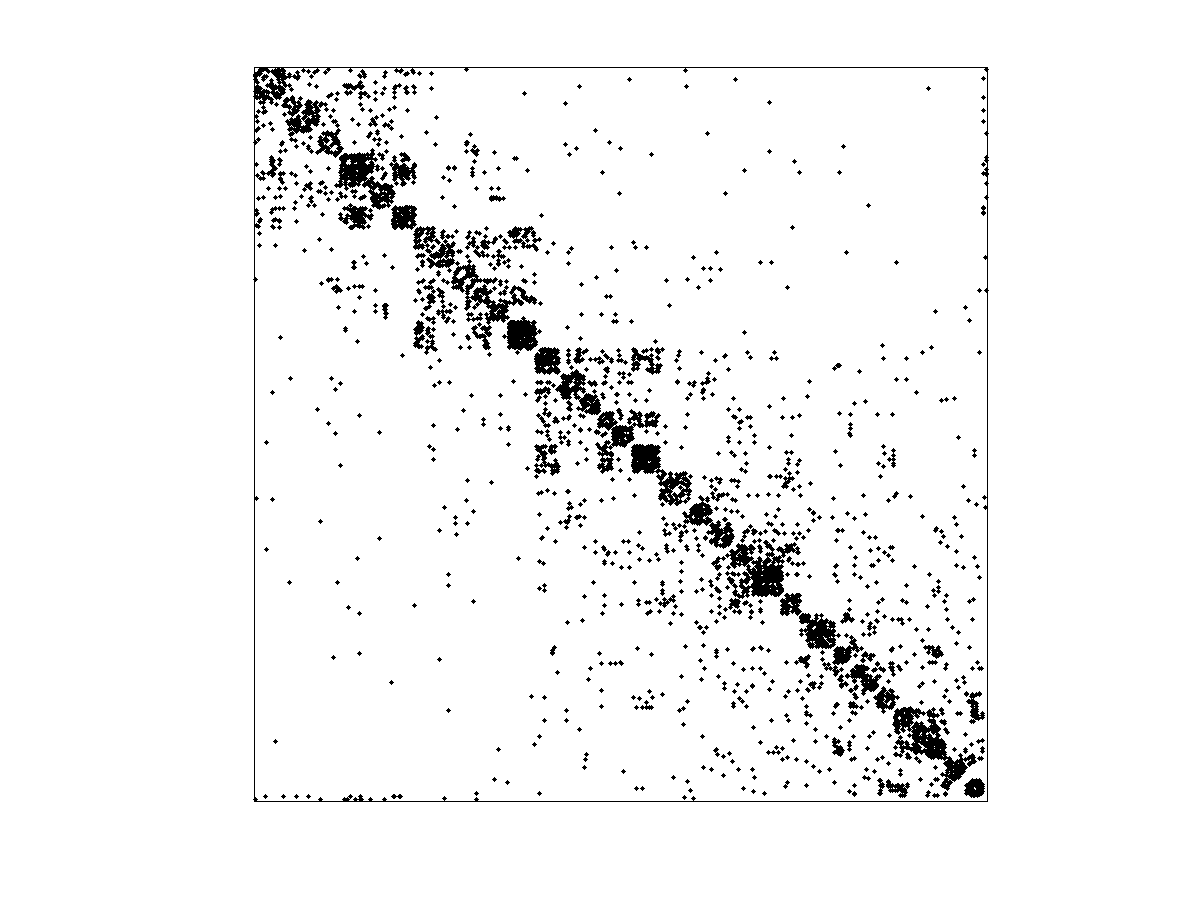}}
}
\subfigure[School 18, $K=5$]{
\makebox{\includegraphics[width=.23\linewidth]{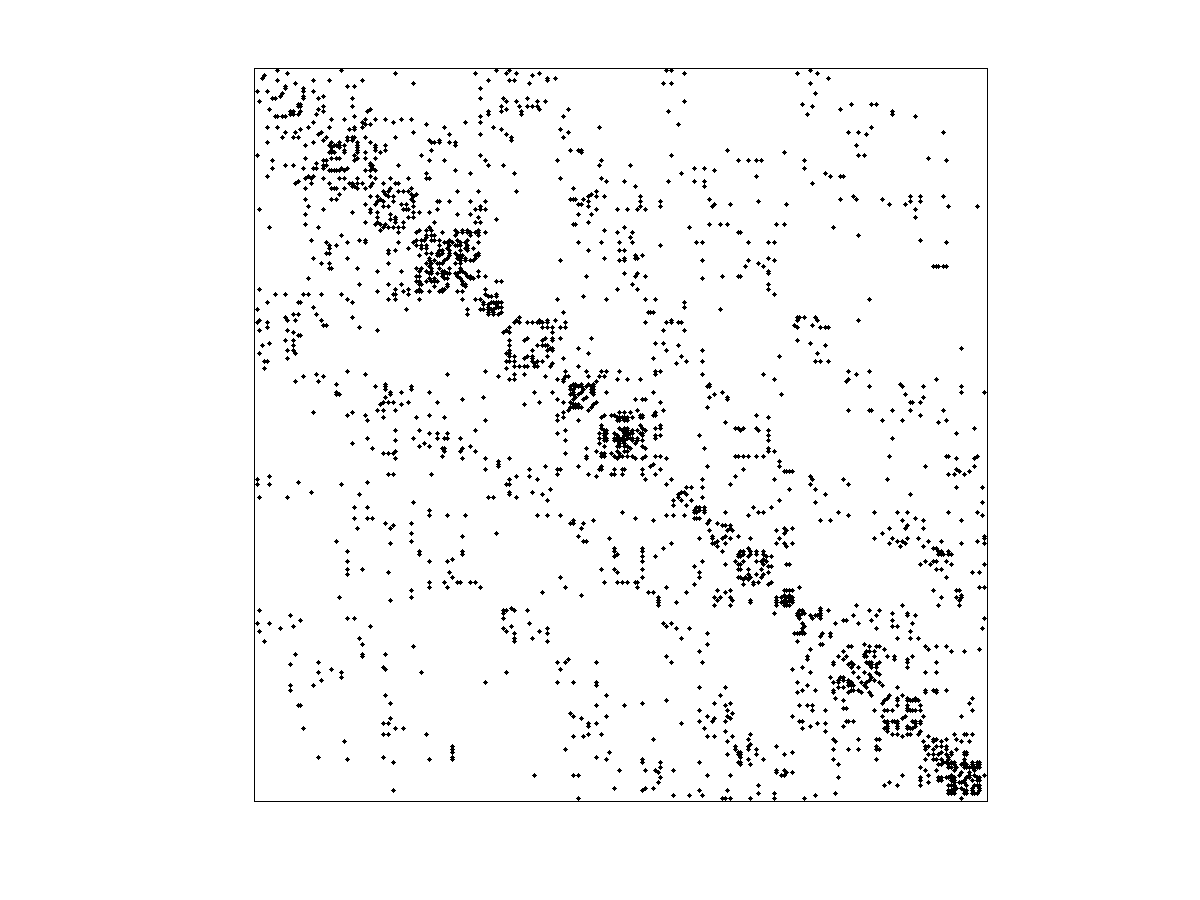}}
}
\subfigure[School 21, $K=6$]{
\makebox{\includegraphics[width=.23\linewidth]{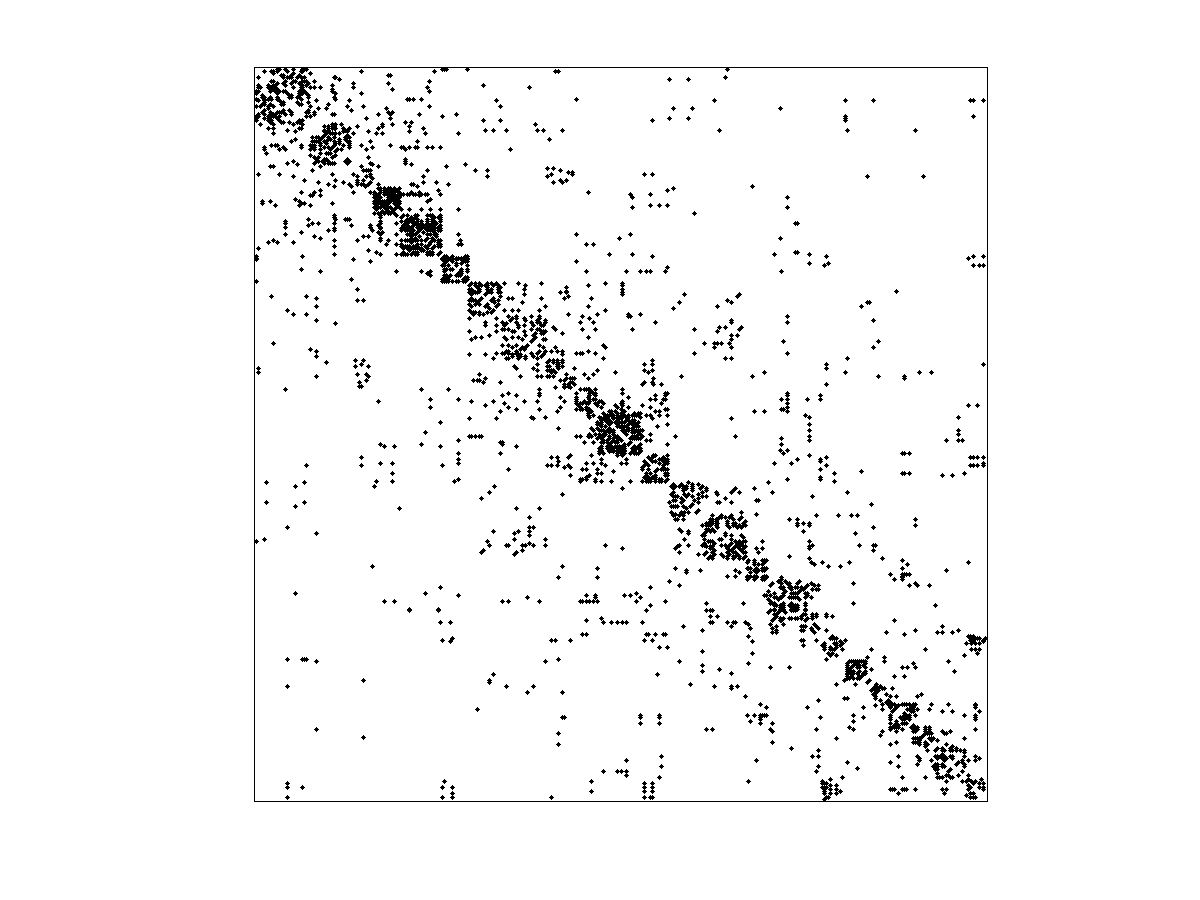}}
}
\subfigure[School 22, $K=5$]{
\makebox{\includegraphics[width=.23\linewidth]{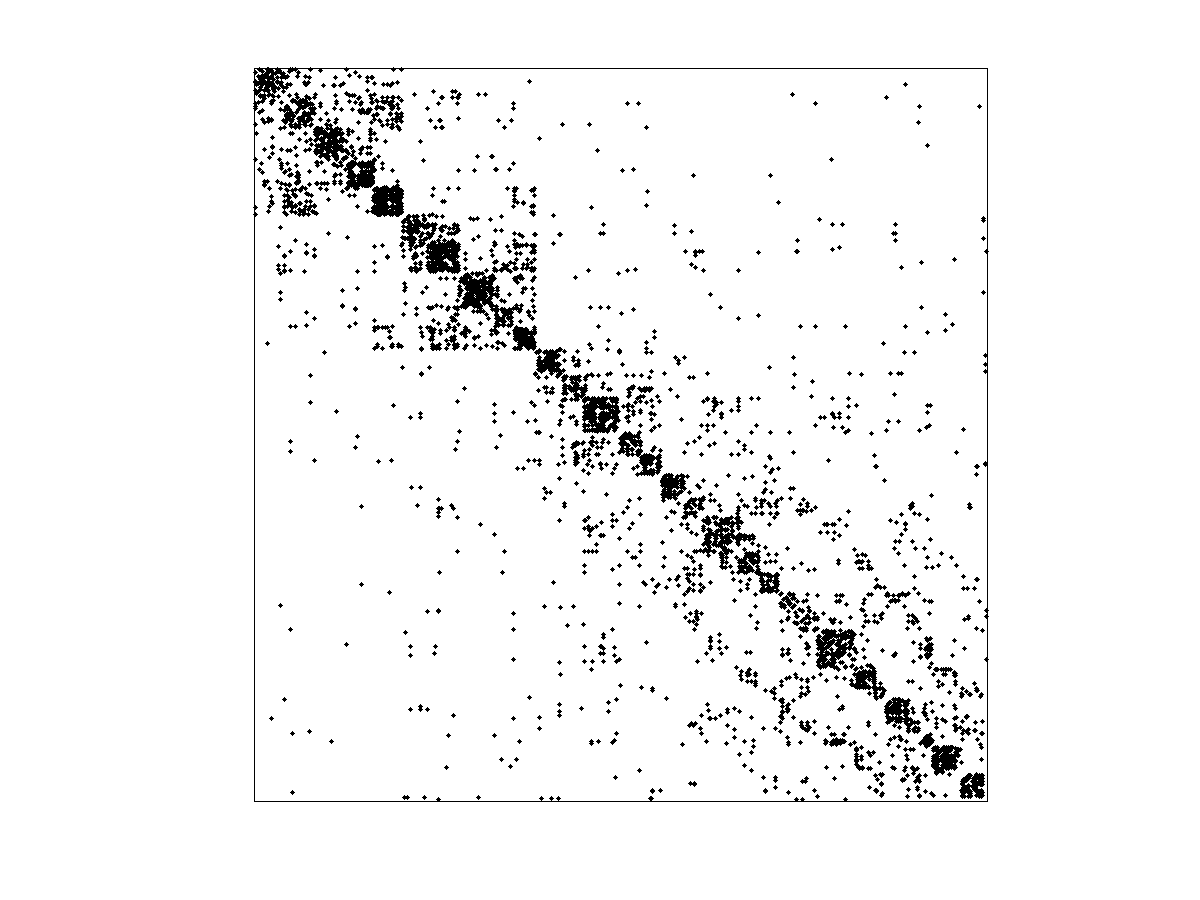}}
}
\subfigure[School 26, $K=3$]{
\makebox{\includegraphics[width=.23\linewidth]{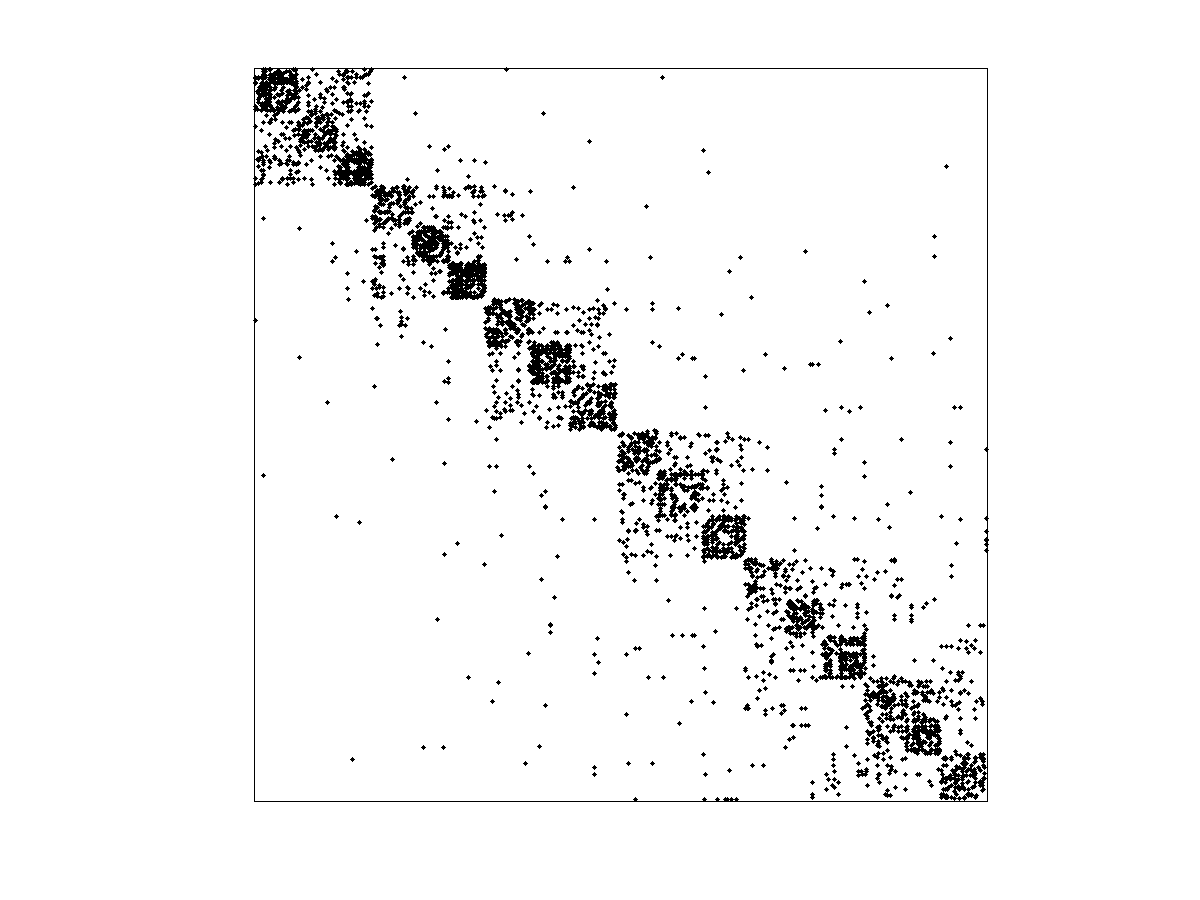}}
}
\subfigure[School 29, $K=6$]{
\makebox{\includegraphics[width=.23\linewidth]{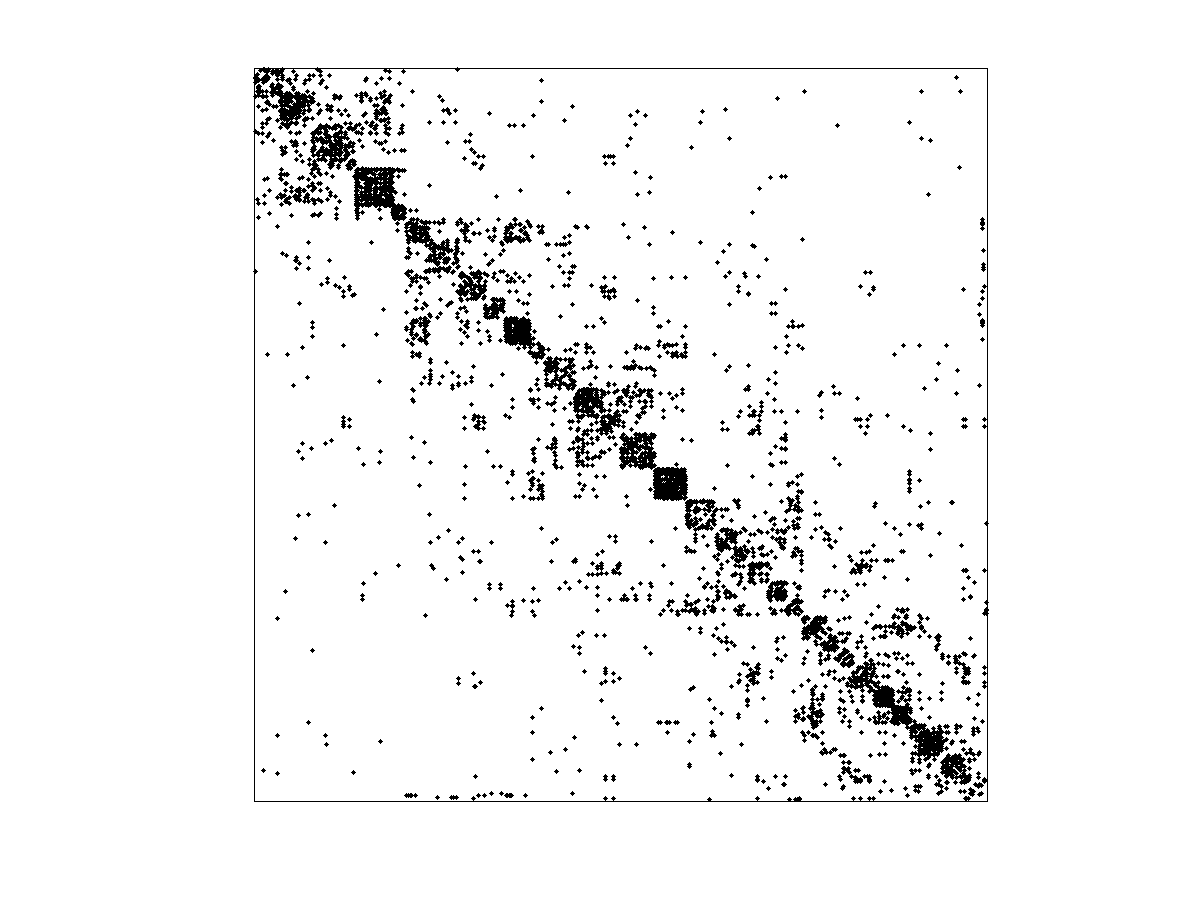}}
}
\subfigure[School 38, $K=5$]{
\makebox{\includegraphics[width=.23\linewidth]{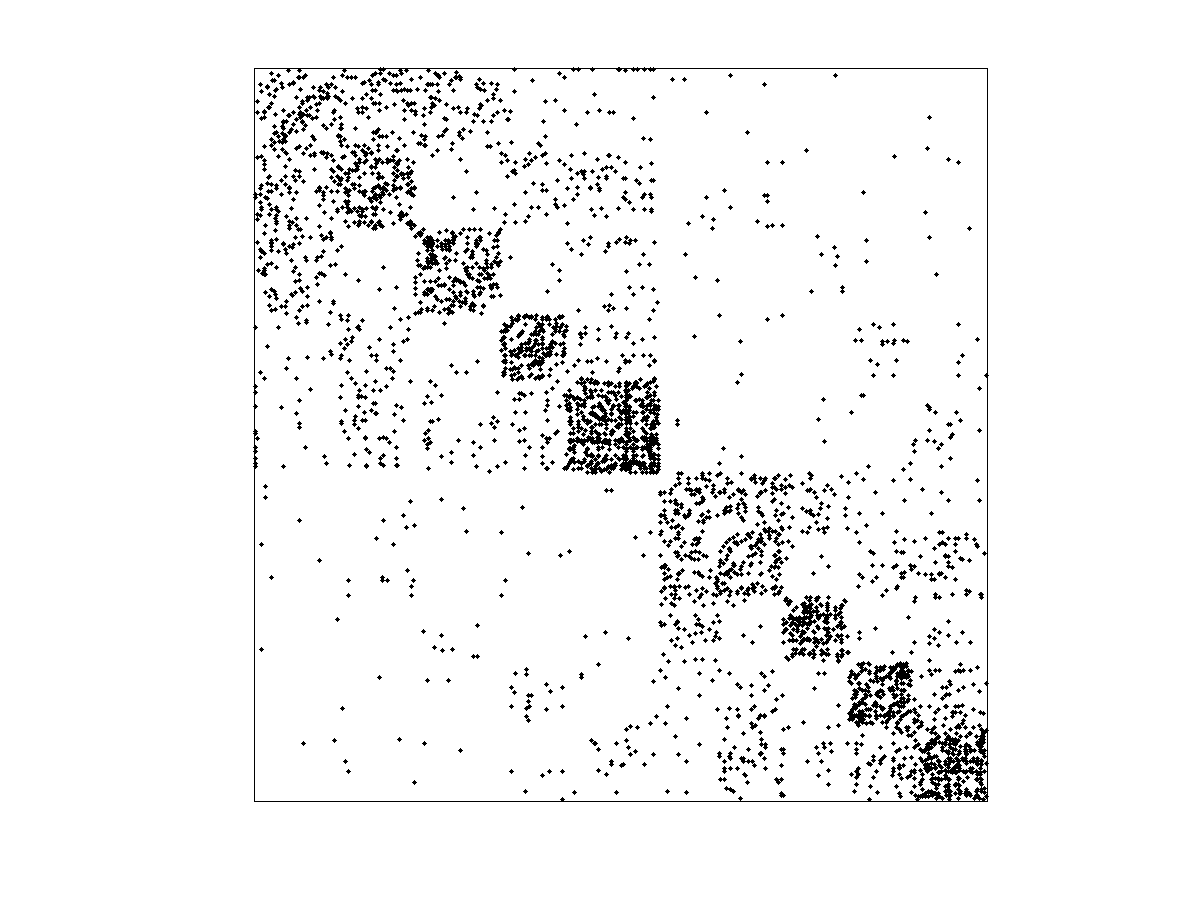}}
}
\subfigure[School 55, $K=4$]{
\makebox{\includegraphics[width=.23\linewidth]{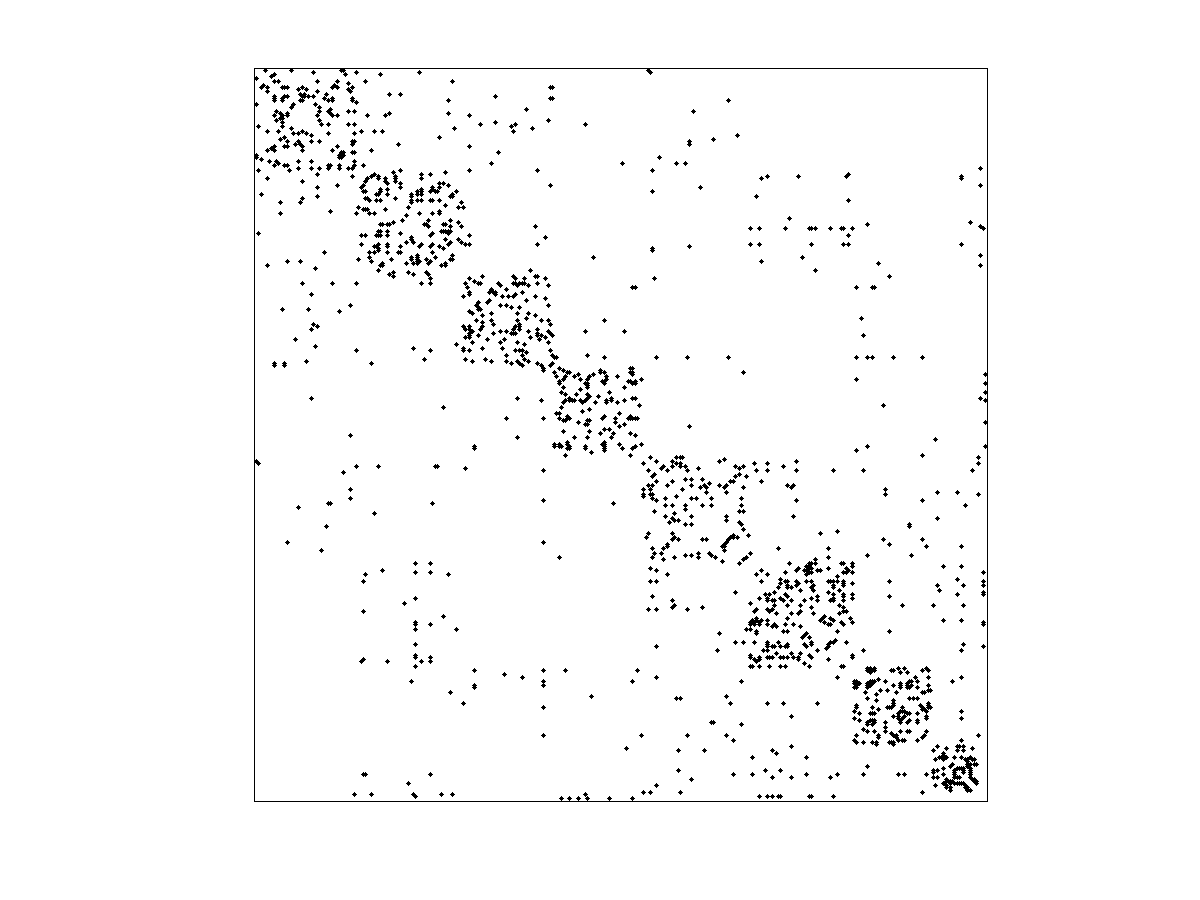}}
}
\subfigure[School 56, $K=6$]{
\makebox{\includegraphics[width=.23\linewidth]{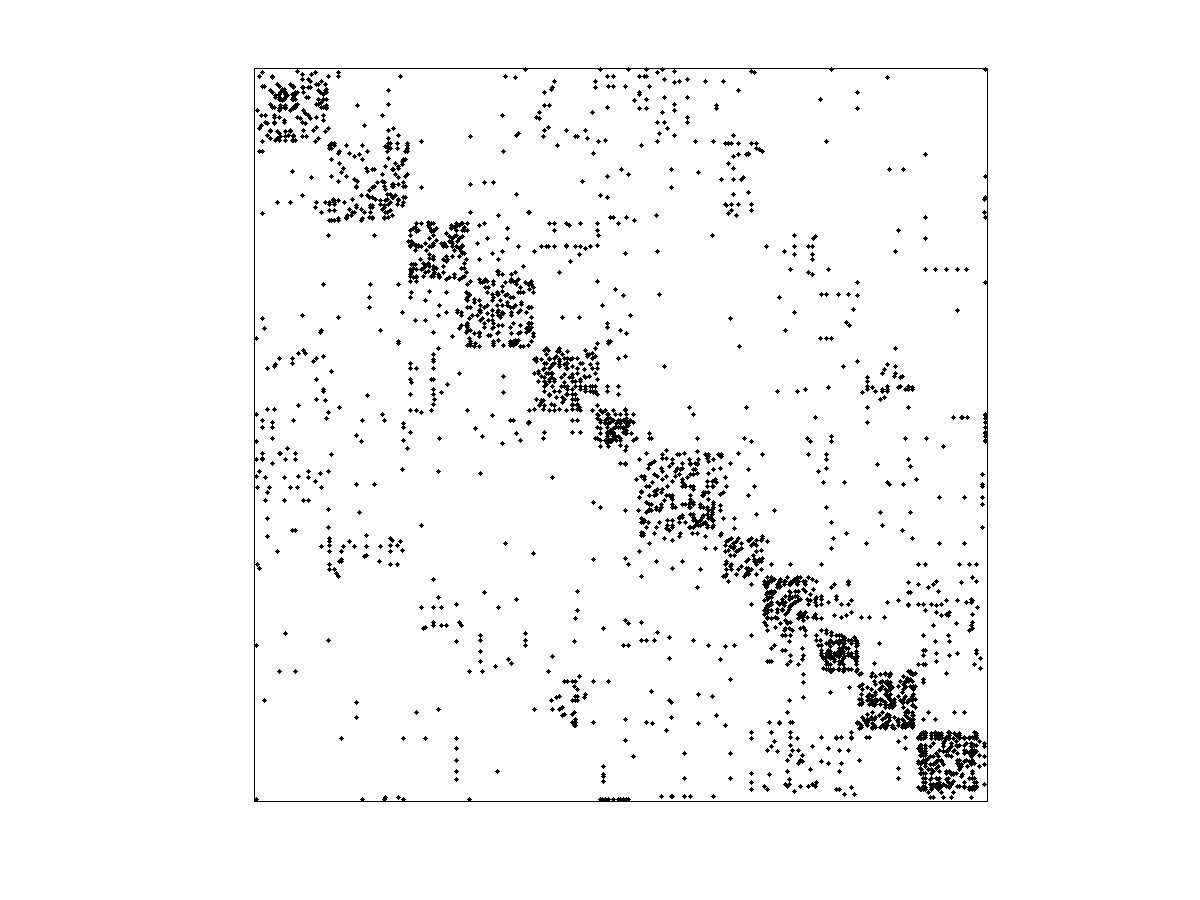}}
}
\subfigure[School 66, $K=6$]{
\makebox{\includegraphics[width=.23\linewidth]{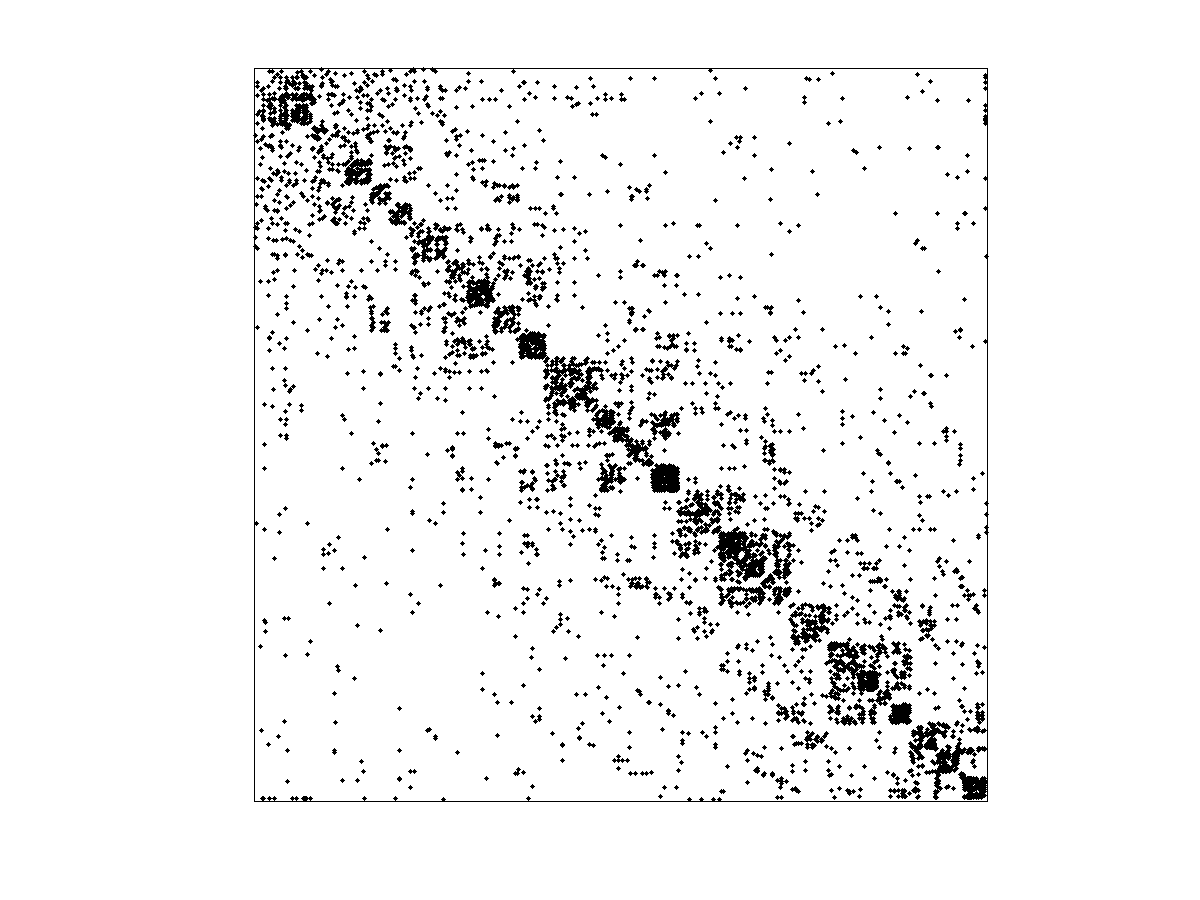}}
}
\subfigure[School 67, $K=3$]{
\makebox{\includegraphics[width=.23\linewidth]{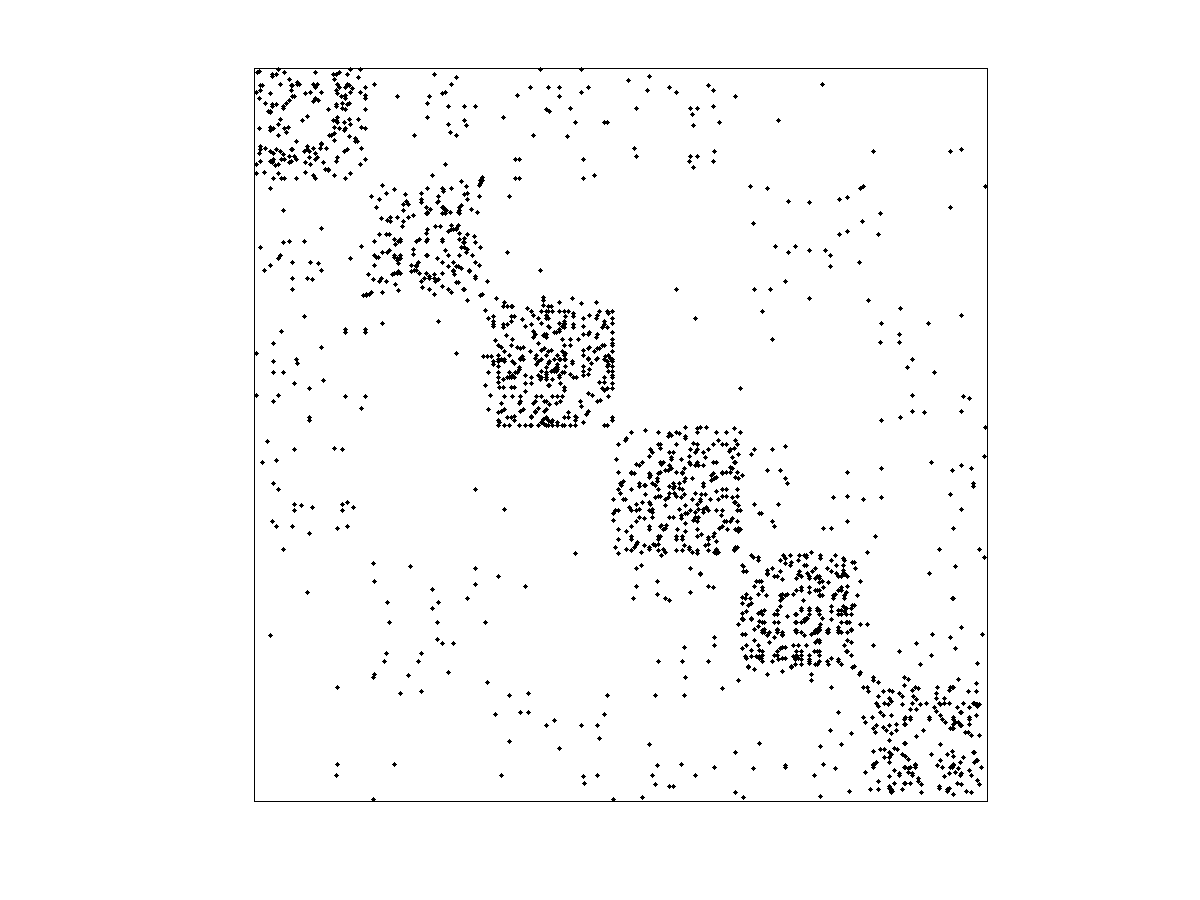}}
}
\subfigure[School 72, $K=4$]{
\makebox{\includegraphics[width=.23\linewidth]{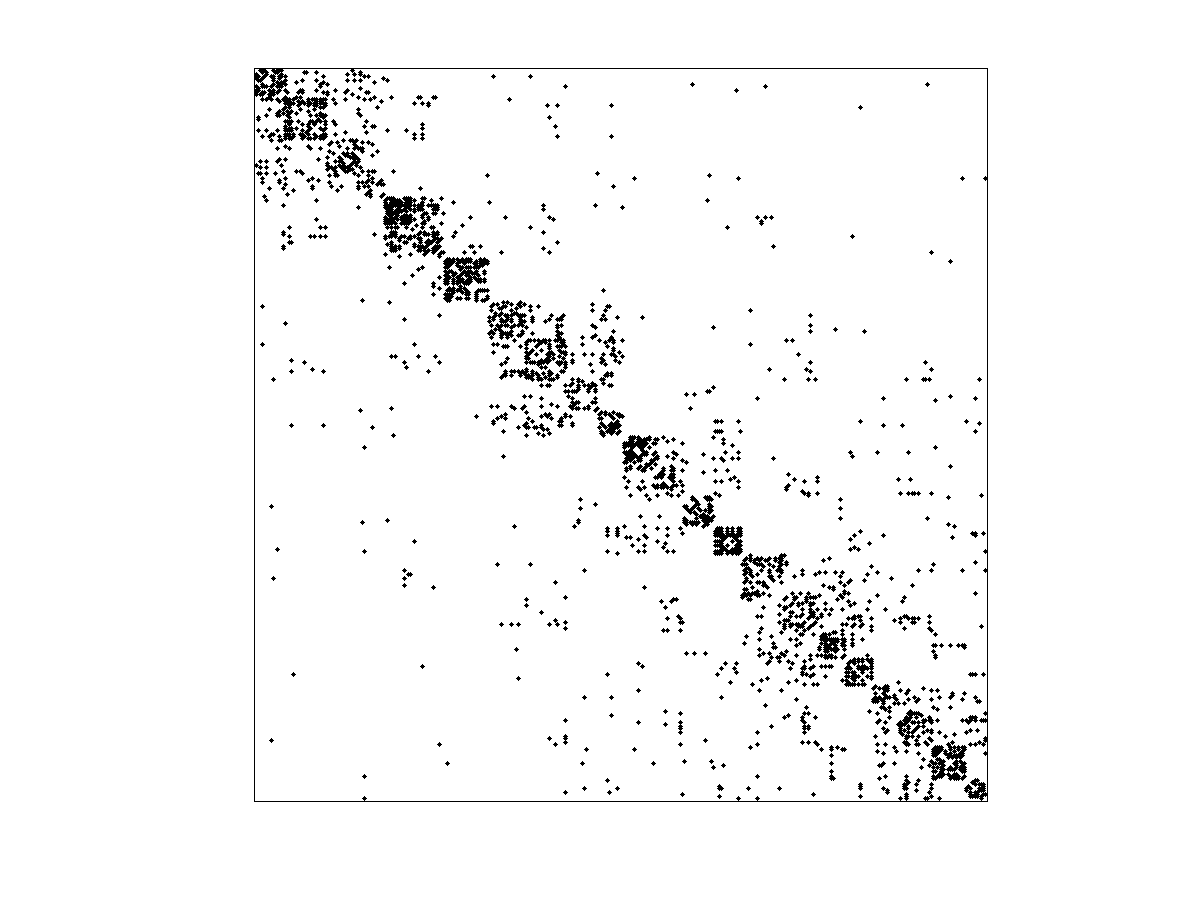}}
}
\subfigure[School 78, $K=6$]{
\makebox{\includegraphics[width=.23\linewidth]{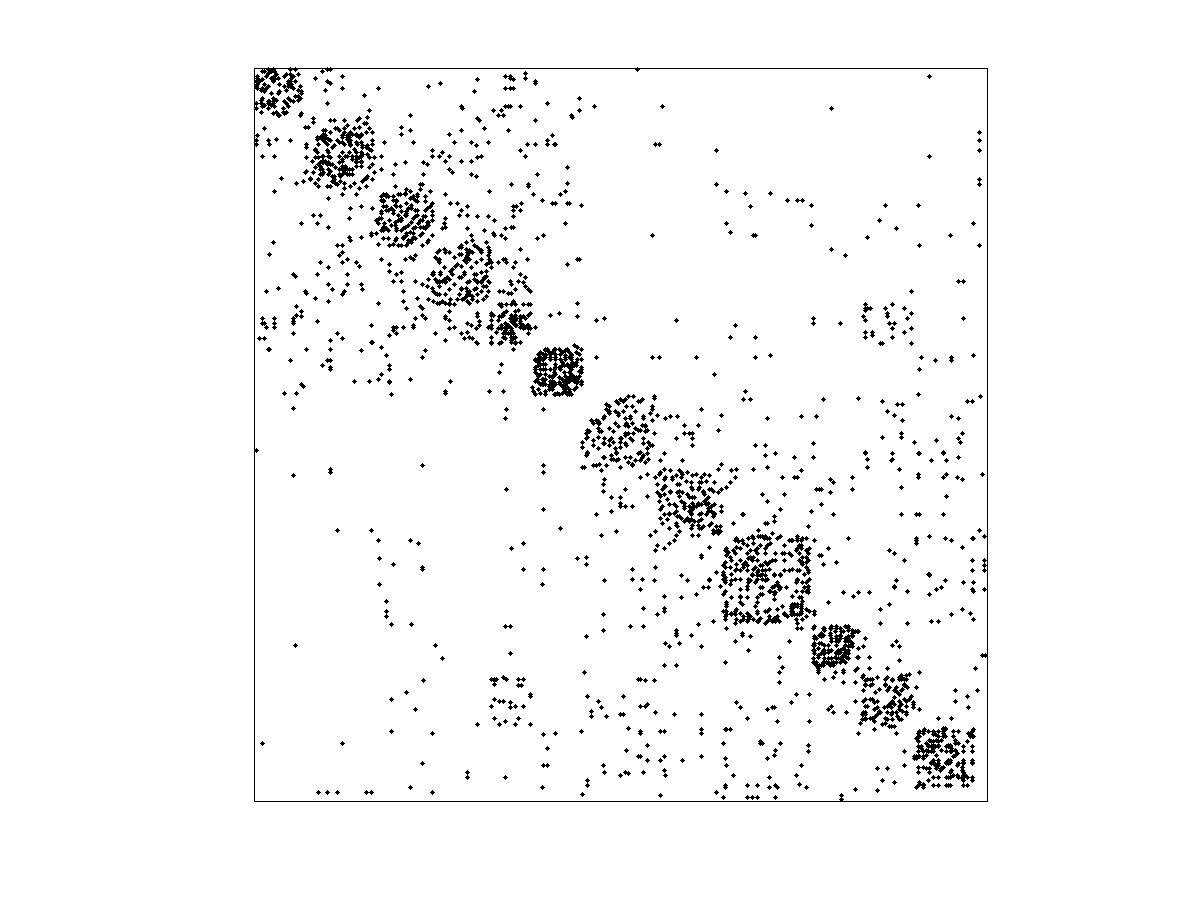}}
}
\subfigure[School 80, $K=4$]{
\makebox{\includegraphics[width=.23\linewidth]{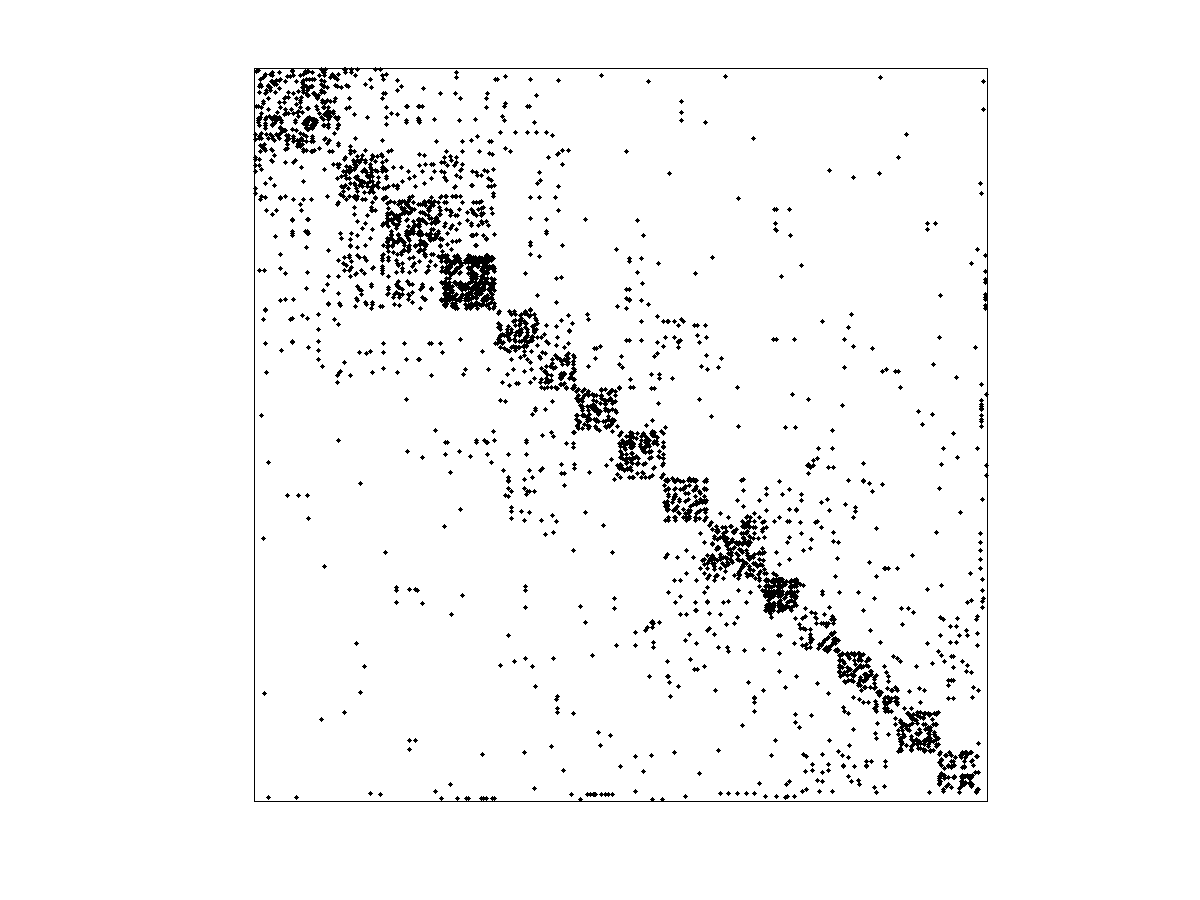}}
}
\caption{Adjacency matrices for schools exhibiting residual block structure as described in Section~\ref{sec:residBlockStruct}, with nodes ordered by grade and corresponding latent classes.}
\label{fig:good-alternate}
\end{figure}

Fig.~\ref{fig:good-alternate} depicts partitions for which the observed $\hat{\Phi}^{(z)}$, fitted for various $K>1$ using the modified version of Algorithm~\ref{alg:1} detailed above,  is ``far'' from its nominal value under the baseline model fitted with $K=1$, in the sense that the corresponding 95\% Bonferroni-corrected confidence set bound is exceeded. We observe that in each partition, the number of apparently visible communities exceeds $K$, and they are comprised of small numbers of students. This effect is due to the intersection of grade and $z$-induced clustering. 

We take as our definition of nominal value the quantity $\hat{\Phi}^{(z)}$ \emph{computed under the baseline model}, which we denote by $\varPhi^{(z)}$.  Table~\ref{tab:Kmult-alternate} lists normalized divergence terms $\binom{N}{2}^{-1} \sum_{a\leq b} n_{ab} D(\hat{\Phi}_{ab}^{(z)} || \varPhi_{ab}^{(z)})$, Bonferroni-corrected 95\% confidence bounds, and measures of alignment between the corresponding partitions $z$ and the explanatory variables. The alignment with the covariates are small, as measured by the Jacaard similarity coefficient and ratio of within-class to total variance\footnote{The alignment scores are defined as follows. The Jacaard similarity coefficient is defined as $|A\cap B| / |A\cup B|$, were $A,B \subset {N \choose 2}$ are the student pairings sharing the same latent class or the same covariate value, respectively. See \citep{traud2010community} for further network-related discussion. Variance ratio denotes the within-class degree variance divided by the total variance, averaged over all classes.}, signifying the residual quality of the partitions, while the relatively large divergence terms signify that the Bonferroni-corrected confidence set bounds for each school have been met or exceeded.

We note that the usage of covariate information was necessary to detect small student groups; without the incorporation of grade effects, we would require a much larger value of $K$ for Algorithm \ref{alg:1} to detect the observed network structure (a concern noted by~\citep{hoff2008modeling} in the absence of covariates), which in turn would inflate the confidence set, leading to an inability to validate the observed structure from that predicted by a baseline model.

\begin{table}
\begin{center}
\begin{tabular}{>{\footnotesize}l|>{\footnotesize}r|>{\footnotesize}r|>{\footnotesize}r|>{\footnotesize}r|>{\footnotesize}r|>{\footnotesize}r|>{\footnotesize}r|}
       &   &                                  &\multicolumn{4}{>{\footnotesize}c|}{Jaccard coefficient / Variance ratio} \\
School & $K$ & Div. (Bound) & Gender & Race & Grade & Degree \\
\hline
10 & 6 & 0.0064 (0.0062) & 0.14 & 0.16 & 0.097 & 0.93 \\
18 & 5 & 0.0150 (0.0150) & 0.17 & 0.19 & 0.14 & 0.88 \\
21 & 6 & 0.0140 (0.0120) & 0.15 & 0.16 & 0.12 & 0.95 \\
22 & 5 & 0.0064 (0.0061) & 0.18 & 0.14 & 0.11 & 0.99 \\
26 & 3 & 0.0049 (0.0045) & 0.25 & 0.21 & 0.13 & 0.99 \\
29 & 6 & 0.0091 (0.0075) & 0.15 & 0.16 & 0.10 & 0.88 \\
38 & 5 & 0.0073 (0.0073) & 0.17 & 0.18 & 0.17 & 0.86 \\
55 & 4 & 0.0100 (0.0100) & 0.20 & 0.18 & 0.21 & 0.97 \\
56 & 6 & 0.0120 (0.0099) & 0.15 & 0.14 & 0.15 & 0.98 \\
66 & 6 & 0.0069 (0.0066) & 0.15 & 0.16 & 0.099 & 0.91 \\
67 & 3 & 0.0055 (0.0055) & 0.25 & 0.23 & 0.25 & 1.00 \\
72 & 4 & 0.0099 (0.0095) & 0.21 & 0.21 & 0.12 & 0.96 \\
78 & 6 & 0.0100 (0.0100) & 0.15 & 0.12 & 0.15 & 0.98 \\
80 & 4 & 0.0054 (0.0053) & 0.20 & 0.19 & 0.15 & 0.99 \\
       \end{tabular}
       \caption{Block structure assessments corresponding to Fig.~\ref{fig:good-alternate}. Small Jacaard coefficient values and variance ratios approaching 1 indicate a lack of alignment with covariates and hence the identification of residual structure in the corresponding partition.}
\label{tab:Kmult-alternate}
\end{center}
\end{table}

\section{Concluding remarks}

In this article we have developed confidence sets for assessing inferred network structure, by leveraging our result derived in~\citet{mypaper}. We explored the use of these confidence sets with an application to the analysis of Adolescent Health survey data comprising friendship networks from 26 schools. 

Our methodology can be summarized as follows. In lieu of a parametric model, we assume dyadic independence with Bernoulli parameters $\{P_{ji}\}$. We introduced a baseline model ($K=1$) that incorporates degree and covariate effects, without block structure. Algorithm~\ref{alg:1} was then used to find highly assortative partitions of students which are also far from partitions induced by the explanatory covariates in the baseline model. Differences in assortativity were quantified by an empirical divergence statistic, which was compared to an upper bound computed from Eq.~\eqref{eq:bound} to check for significance and to generate confidence sets for $\{P_{ji}\}$. While the upper bound in Eq.~\eqref{eq:bound} is known to be loose, simulation results in Figure \ref{fig:divergence histogram} suggest that the slack is moderate, leading to useful confidence sets in practice. 

In our procedure, we cannot quantify the uncertainty associated with the estimated baseline model, since the nodal parameter estimates lack consistency. As a result, we cannot conduct a formal hypothesis test for  $\Theta=0$. However, for a baseline model where the MLE is known to be consistent, we conjecture that such a hypothesis test should be possible by incorporating the confidence set associated with the MLE.

Despite concerns regarding estimator consistency in this and other latent variable models, we were able to show that the notion of confidence sets may instead be used to provide a (conservative) measure of residual block structure.  We note that many open questions remain, and are hopeful that this analysis may help to shed light on some important current issues facing practitioners and theorists alike in statistical network analysis.

\subsection*{Acknowledgments}

Work supported in part by the National Science Foundation, National Institutes of Health, Army Research Office, and Office of Naval Research, USA.  Additional funding was provided by the Harvard Medical School's Milton Fund.
This research uses data from Add Health, a project funded by the Eunice Kennedy Shriver National Institute of Child Health and Human Development. Information on how to obtain the Add Health data files is available online at http://www.cpc.unc.edu/addhealth.
We thank the reviewers for their comments and suggestions that helped improve the paper.


\bibliographystyle{plainnat}

\end{document}